# Electrically driven random lasing from a modified Fabry-Perot laser diode

Antonio Consoli[1,2]*, Niccolò Caselli[1†#], Cefe López[1§]

Random lasers (RLs) are intriguing devices with promising applications as light sources for imaging, sensing, super resolution spectral analysis or complex networks engineering. RLs can be obtained from optically pumped dyes, optical fibers and crystals, or electrically pumped semiconductor heterostructures. Semiconductor RLs are usually fabricated by introducing scattering defects into the active layer, adding a degree of complexity to the fabrication process and losing the ease of realization potentially offered by disordered structures. Ready availability of electrically pumped RLs, avoiding costly fabrication approach, would boost the use of these devices in research and applications. Here, we realize an incoherent semiconductor RL by simply processing the output mirror of an off-the-shelf Fabry-Perot laser diode via controlled laser ablation. Optical feedback provided by the intact back mirror and the ablated front mirror results in multi-mode random lasing with low spatial coherence and speckled output emission profile.

## Introduction

In random lasers (RLs) optical feedback for lasing action is provided by scattering elements usually embedded in the active material[1]. Due to their peculiar characteristics, RLs are promising devices for different applications, e.g. laser-based speckle free imaging[2], sensors[3], spectral measurement with super resolution[4] and networks of coupled resonators[5].

The first proposal of a "non-resonant" laser, i.e. a laser cavity based on optical feedback from scattered light, consisted of a gain material placed between one mirror and one diffuser[6]. The same authors proposed an alternative implementations of this concept[7], including a distributed feedback architecture, where the refractive index of the active material randomly varied in space. This approach has been extensively used, leading to the first demonstrations of random lasing[8–10], and it still is the most widespread fabrication method.

RLs with distributed feedback have been obtained in a large range of materials[11–18], basically mixing any available scattering element and active medium. For instance, dyes, in solid and liquid forms, have been largely exploited for the realization of RLs, with the stringent requirement of pulsed optical pumping, which strongly limits the ease of operation and applications. Random fiber lasers, in which distributed Rayleigh scattering provides the required feedback, are pumped by laser diodes coupled to the fibers, resulting in high power CW output[19]. Electrically pumped semiconductor RLs[20] are indeed suitable devices both for research and application due to their reliability, ease of operation, small footprint and potentially low-cost fabrication on large volumes.

Semiconductor RLs can be fabricated by introducing a matrix of defects inside the active layer, where the size, shape and spatial distribution of the defects are designed so that the introduced disorder is controlled and tailored *ad hoc*. This approach has been used for the realization of quantum cascade RLs[21–24] in which holes are etched into the active layer with precise sizes and positions. Alternatively, truly disordered structures, i.e. with unpredictable sizes, shapes and positions, can also be used, by deposition of polycrystalline film with grains[25–30], by active material growth from randomly distributed seeds on substrate[31–34] or by introduction of random defects into the active layer[35–37].

In all the previously mentioned devices the material growth process must include a specific fabrication step for introducing a random variation of the refractive index into the active layer. This requirement is in contrast with the typical ease of fabrication of RLs, e.g. in devices consisting of scattering powders and dyes, which allow manageable processing but require optical pumping. Simple realization of disordered structure in a semiconductor material that could allow electrically pumped random lasing would be cornerstone for building future applications of RLs.

Our group has previously demonstrated that random lasing can be efficiently achieved in a device where defects are placed at the edges of the active material, instead of being distributed inside of it[38]. In an analogy with Fabry-Perot (FP) lasers, we substituted mirrors for two differently disordered scattering elements which provide feedback and output coupling. We observed the same spectral random lasing signature, specific of each device considered, emitted from both laser ends. Therefore, the existence of a single optical resonator, consisting of the pumped active region and the two passive scattering elements with different disorder, was demonstrated. We previously realized such devices with liquid or solid dye and defects at edges consisting of scattering nanoparticle aggregates, natural raw materials or ablated polymer surfaces[39,40]. A review on architectures based on distributed/non-distributed feedback is given in Supplementary Material, Section 1.

[1]Instituto de Ciencia de Materiales de Madrid (ICMM), Consejo Superior de Investigaciones Científicas (CSIC), Calle Sor Juana Inés de la Cruz, 3, 28049 Madrid, Spain. [2]ETSI de Telecomunicación, Universidad Rey Juan Carlos, Calle Tulipán, 28933 Madrid, Spain [†]Current address: Dep. Química Física, Universidad Complutense de Madrid, Avenida Complutense, 28040 Madrid, Spain. *e-mail: antonio.consoli@urjc.es, #e-mail: ncaselli@ucm.es, §email: c.lopez@csic.es.

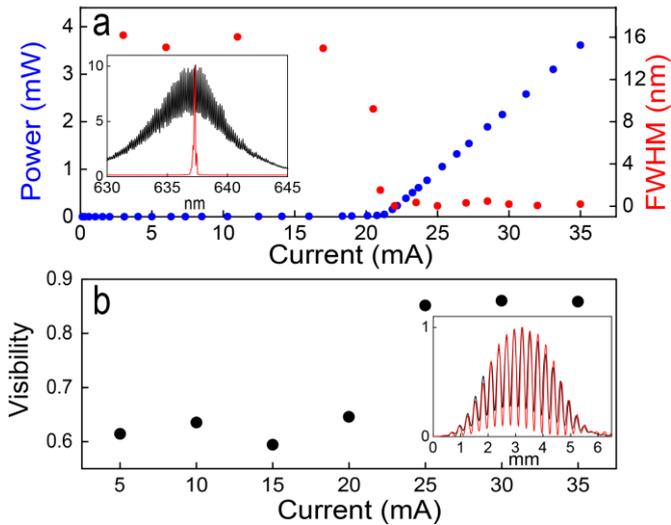

**Figure 1. | Characterization of the original FP laser diode.** (a) Power output (blue dots, left axis) and spectral at full-width at half-maximum (FWHM, red dots, right axis) versus injected current; inset: normalized emission spectrum below threshold, at 3 mA (black line) and above threshold, at 30 mA (red line). (b) Fringe visibility γ, as defined in the text, as a function of injected current; inset: interference fringes measured with the two slits set-up, for currents of 5 mA (black curve) and 35 mA (red curve).

Here, we demonstrate that architectures made of resonator with disordered mirrors can be used for obtaining a random laser diode by employing a simple modification process of commercially available, low cost, FP laser diodes. We use a pulsed, high energy laser beam for surface ablation of the front mirror of the FP laser diode which, through roughness, induces defects with random shapes and sizes on the emitting area of the active layer. The back mirror of the laser diode is left unchanged, in a configuration conceptually similar to the first proposal of a "non-resonant" laser[6]. We characterize the output power, spectral and spatial emission and spatial coherence of the obtained device and compare it to the original FP laser diode, obtaining incoherent random lasing emission that exhibit promising performances for future applications.

## Fabry-Perot Laser Diode Emission

The device under study is an AlGaInP laser diode with multi quantum well structure in a TO can package.

The laser output is collimated by a 100× microscope objective (NA = 0.9), and power and spectral measurements are performed with a calibrated photodetector and with a spectrometer, respectively, see Methods for more details.

Lasing threshold is found to be 21 mA, the slope efficiency is about 0.27 W/A and the spectral width drops from 16 nm below threshold to 0.3 nm above threshold, see Fig. 1a. Typical emission spectra of FP lasers below and above threshold, 20.7 mA and 30 mA, black and red lines, respectively are shown in Fig. 1a inset. Below threshold, spontaneous emission is dominant with respect to stimulated emission and the gain profile is modulated by the equally spaced longitudinal modes of the FP cavity. Above threshold, modal competition at the gain centre wavelength produces an almost single mode spectrum with a dominant peak and few side modes.

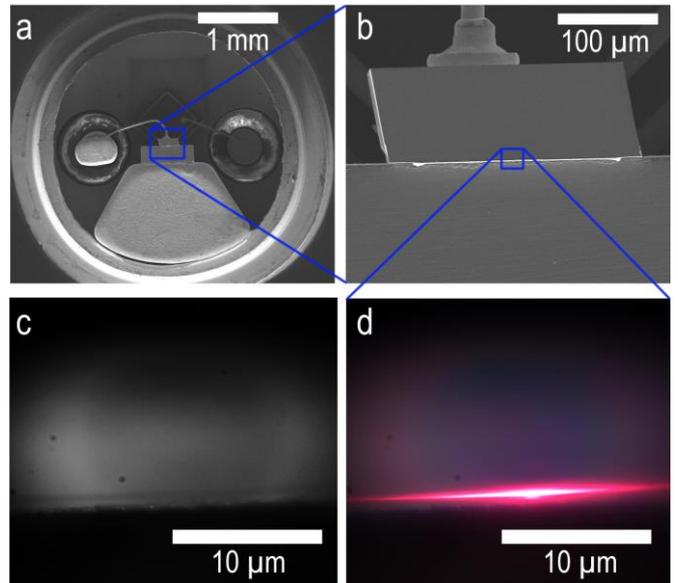

**Figure 2. | The original laser diode.** (a) SEM image of the opened can, the laser dice is in the center, placed on the silicon substrate and the heatsink, scale bar is 1 mm. (b) SEM image of the laser dice, the anode contact on the top and the silicon substrate on the bottom of the image, scale bar is 100 μm. Optical microscope image of the dice, when 0 mA (c) and 3 mA (d) are applied. The active region is the dark gray stripe, near the silicon substrate (black region at the bottom of the image).

Spatial coherence of the FP laser emission is measured by shining the collimated beam onto a black screen through two parallel slits, 50 μm wide and separated by 500 μm. Radiation from the slits is measured by a CCD camera at different injection currents and detected intensities are summed up in CCD pixel columns parallel to the slits. Fringes appear due to the interference between the wavefronts exiting each slit, allowing us to estimate the transverse spatial coherence of the laser from the visibility $\gamma = (I_{MAX} - I_{MIN})/(I_{MAX} + I_{MIN})$, where $I_{MAX}$ and $I_{MIN}$ are the peak and trough intensities of the interference fringes, at the central peak, respectively. In Fig. 1b, we show γ as a function of the injected current. An average value of γ = 0.62 is found below threshold (i.e. for currents up to 20 mA). Above threshold, the visibility steeply increases to an average value of 0.85, for currents between 25 mA and 35 mA. Fringes measured below (5 mA) and above threshold (35 mA) are shown in the inset of Fig. 1b.

## Front mirror modification

The can of the laser diode is removed for granting access to the chip and the laser output mirror. The laser dice is mounted on a silicon substrate placed on the heat sink and cathode terminal, as shown in Fig. 2a, where the monitoring photodiode can also be seen on the background. The active layer is located at the bottom of the dice, close to the silicon substrate, as verified by imaging the emission injecting a small amount of current, see Fig. 2d. The black area at the bottom in Fig. 2c,d is the silicon substrate and the thin active region can be seen close to the edge of the dice as a dark gray stripe when no current is applied, Fig. 2c, and where a bright red emission is observed when applying 3 mA, Fig. 2d.

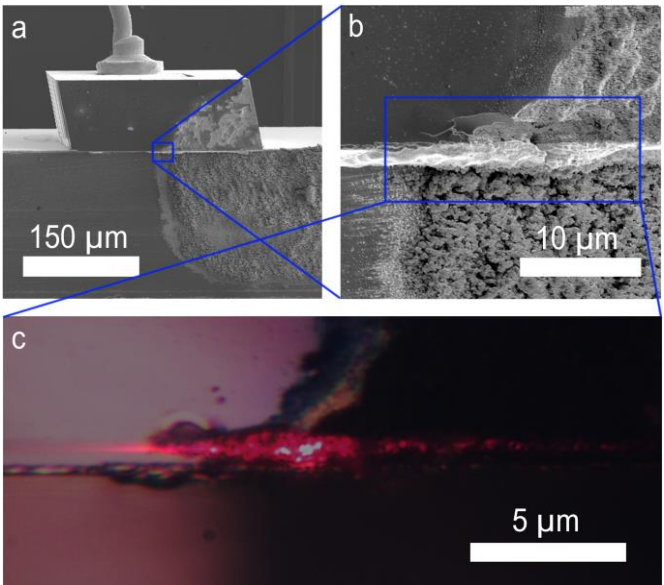

**Figure 3. | The modified laser diode.** (a) SEM image of the laser dice on the silicon substrate, the corrugated surface corresponds to the area hit by the ablating laser pulses, which extends to the laser dice and to the silicon substrate. (b) SEM image of the emitting area of the active region, including part of the dice (upper half of the image) and of the silicon substrate (lower half of the image). (c) Optical microscope image of the emitting area of the active region when a current of 16 mA is injected: dark areas correspond to the ablated regions, which strongly scatter the incident white light used for illumination.

The laser source used for surface ablation is a high energy femtosecond laser (Coherent Libra) emitting at 800 nm, with 100 fs long pulses at a repetition rate of 1 kHz, see the Methods section for more details on the ablation procedure.

Results obtained after blasting the front mirror of the laser diode with 200 pulses at 100 µJ/pulse average energy are shown in Fig. 3. The beam spot size is about 350 µm in diameter, overlapping with the emitting area and avoiding the top electrode. The induced surface roughness is estimated to be in the submicron range, according to horizontal (0.52 µm) and vertical (0.15 µm) correlation lengths derived from the gray scale sections of SEM images of the active region modified by processing (see Section 2 in Supplementary Material for more details). The modified laser diode shows intense emission when current is injected, characterized by bright and dark spots randomly distributed along the active layer, as shown in Fig. 3c.

We remark that the use of pulsed laser ablation is motivated by considering that other surface modification processes, e.g. chemical etching or nano-powders depositions, would produce longer autocorrelation lengths or lower index contrast, respectively[41,42].

## Random Laser Diode Emission

The output power of the modified laser diode and its emission linewidth are plotted in Fig. 4a as a function of the injected current.

Output power increases with current and lasing threshold is found at about 32 mA. The efficiency slope, 1.4 mW/A remarkably decreased with respect to the original FP laser diode (0.27 W/A). This drastic performance loss can be attributed to the degradation of the mirror upon ablation. Although detrimental for the device regarding output power, the induced losses allow for generating sufficient scattering feedback for random lasing action and are key to change the character of the emission.

The modified device's emission full-width at half-maximum (FWHM) decreases with current from 16.8 nm, at 10 mA, to 7.6 nm, at 55 mA, see Fig. 4a. This spectral narrowing above threshold is the fingerprint of lasing action.

We now proceed to study the spectral and spatial properties of the emission of the modified laser diode. The far-field angular emission[43] is characterized with the hyperspectral imaging set-up described in Methods. Briefly, a multimode optical fibre is scanned across the back-focal plane of the collecting objective, so that the spatial coordinates of the fibre tip correspond to azimuthal angles of emission from the edge of the laser diode. In Fig. 4b, we show the angular intensity distribution of the modified laser emission integrated over the entire wavelength range of detection, for an injected current of 50 mA. We collect a large half angle of emission of about 45° with respect to the axis of the pristine laser, observing that most of the radiation falls within 25°. The obtained profile is characterized by an irregular pattern of maxima and minima of intensity, typical of RL modes[44], produced by the particular corrugations performed on the front mirror by the ablation process.

We measure the spectrum as a function of current at a fixed point in the far-field, indicated by the crosshair in the intensity map in Fig. 4b, located in a high intensity area of emission. Measured spectra are plotted in Fig. 4c. We observe a spectral profile consisting of a large gaussian background with superimposed narrow peaks, exhibiting sub-nanometer linewidths and without wavelength periodicity. As a function of the injected current the amplitude of each peak increases while its resonant wavelength remains unaltered. Therefore, we identify these peaks with the lasing modes of the new cavity formed between the disordered front mirror and the flat back mirror.

Spectra acquired at other angles (by changing the collection fibre position in the back-focal plane) present the same characteristic gaussian background, with randomly distributed narrow peaks on top, that show constant wavelength positions independent of injected current. Peaks vary in number and position, for different detection direction, thus varying the spectral signature. The map of angular intensity distribution for a given mode can be obtained by extracting the intensity integrated in narrow windows (0.25 nm) around the associated peak from the acquired spectra and plotting it as a function of the corresponding angles. This is equivalent to setting the mono-chromator to a fixed wavelength and scanning the collection fiber. Examples of such a detailed hyper spectral characterization is given in Section 3 of Supplementary Material. We attribute this behavior to an emission characterized by a large number of lasing modes with strong mutual interaction in space and frequency, as previously observed in incoherent RLs[45,46].

The spectrum shown in Fig. 4d is the total emission obtained by summing up, for each wavelength, all the collected spectra on the far-field image of Fig. 4b. A large gaussian background with FWHM of 7.8 nm is observed, typical of incoherent random lasing emission. In fact, narrow peaks from all collected spectra, whose wavelength and number randomly vary from point to point, sum up averaging the full spectral emission into a gaussian shape. The sharp single mode emission with sub-nanometer linewidth of the original FP laser diode, observed above threshold in the inset of Fig. 1a, is lost after introducing disorder in the laser diode front mirror, transforming the single mode FP laser into a multimode RL.

A residual periodicity of about 0.16 nm is still observed at the spectrum peak where some small intensity ripples are found, see the inset in Fig. 4d. We investigate the peak periodicity through the intensity Fourier transform (IFT) of normalized spectra and report it as a function of equivalent cavity length $L$, where $L = \lambda_0^2/(2\,\Delta\lambda\,n)$, with $\lambda_0$ the central wavelength, $\Delta\lambda$ the wavelength period and $n = 3.5$ the refractive index. For the original laser diode, the IFT obtained below threshold (20.7 mA) shows a peak over a noisy background at $L = 365$ µm, corresponding to $\Delta\lambda = 0.16$ nm, in agreement with the actual cavity length observed by the electronic microscope. For the modified laser diode, the IFT of the single spectrum collected at a fixed point of the far-field shows no clear fingerprint of periodicity. While the IFT of the sum of all spectra shows, lying over a noisy background, a peak coinciding with the original laser diode cavity length, but with an intensity decreased of about 50 dB. This is attributed to the degree of disordered introduced onto the modified mirror, as confirmed by numerical simulations presented in the Theoretical model section and in Section 4 of Supplementary Material.

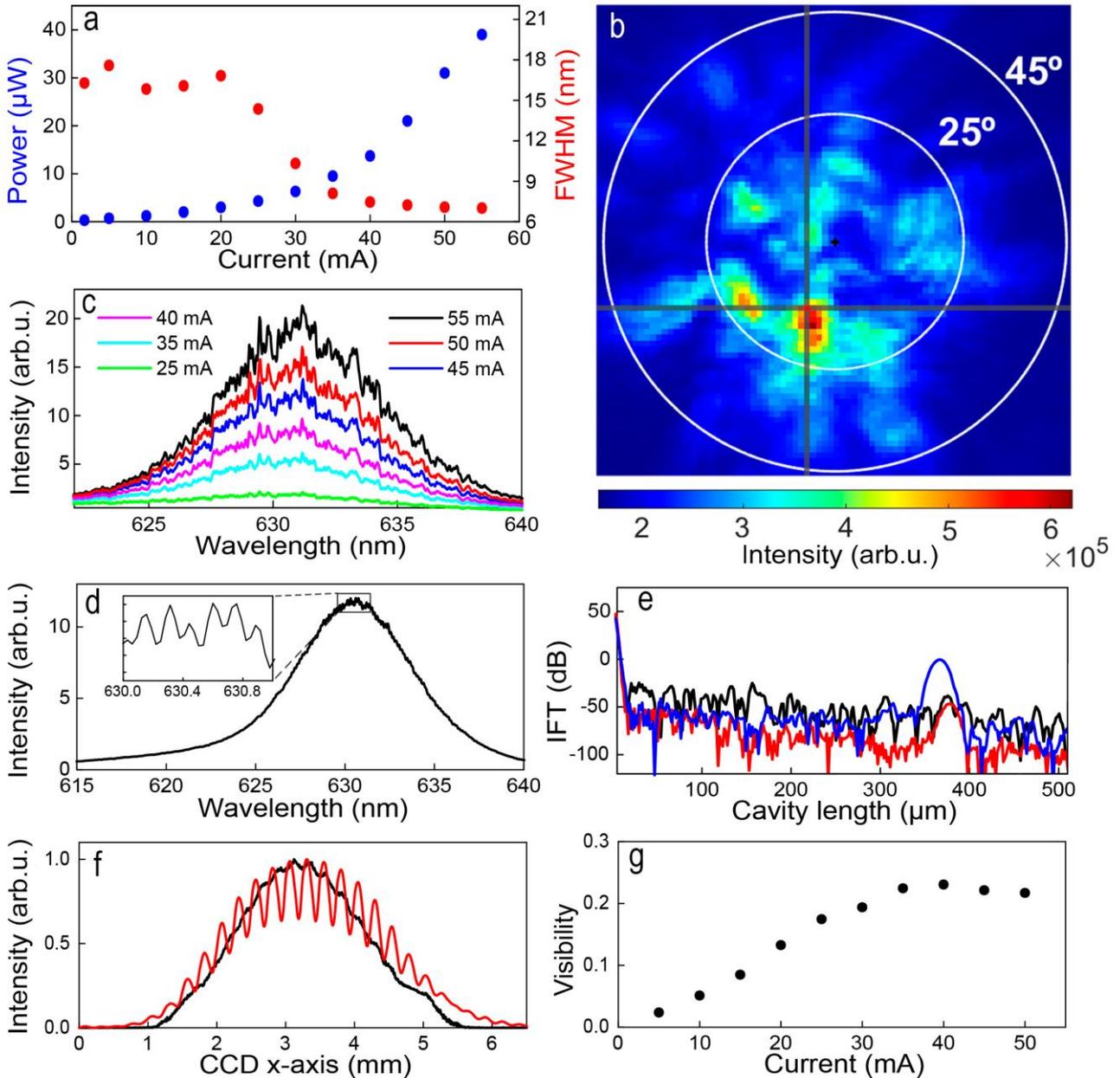

**Figure 4. | Modified FP laser diode emission.** (a) Output power (blue dots, left axis) and FWHM (red dots, right axis) as a function of current. (b) Angular intensity distribution at injection current of 50 mA, obtained by integrating the spectrum collected at each point of the back-focal plane over the entire wavelength detection range, the pixel in the crosshair indicates the point of collection for spectra shown in panel (c). (c) Spectra acquired at given emission angle, for current ranging between 5 mA and 50 mA. (d) Full emission spectrum obtained as the sum of all the collected spectra on the back-focal plane, for a pumping current of 50 mA. The inset shows a detailed view of the curve peak. (e) Intensity Fourier Transform (IFT) as a function of the equivalent cavity length of the spectrum acquired from the original laser diode at 20.7 mA (blue curve) and from the modified laser diode at 50 mA evaluated at a fixed emission angle selected in (b)-(c) (black line) and integrated over all the acquired angles (red line). (f) Interference fringes measured with the two slits set-up, for 5 mA (black curve) and 50 mA (red curve). (g) Fringe visibility γ, as defined in the text, as a function of current.

The spatial coherence of the modified laser diode emission is measured with the same set-up used for the original FP laser diode. In Fig. 4f, we show the fringes obtained for injected current values of 5 mA and 50 mA. A very low fringe contrast is observed for 5 mA which increases with current. This behavior is quantified by calculating the visibility γ, shown in Fig. 4g. Below threshold, it varies between 0.02 at 5 mA and 0.19 at 30 mA. Above threshold, γ remains almost constant at an average value of 0.22. Compared to the lowest value (γ = 0.6) below threshold and the highest value (γ = 0.85) above threshold in the original FP laser, the visibility, hence spatial coherence, of the modified device has decreased abruptly. We attribute the strong reduction of visibility to an equivalent reduction in spatial coherence due to the participation of a very large number of lasing modes, as observed in degenerate cavities and RLs[45].

## Theoretical model

The presented experimental results are benchmarked against the theoretical framework of RLs with non-distributed feedback[38], in which phase and amplitude of disordered reflectors are assumed to vary randomly with frequency in a sub-nanometric scale. This assumption shifts the spatial complexity of the device to the spectral domain, where light at each frequency in the gain bandwidth, after being scattered back from the disordered mirrors, re-enters the cavity with random phase and amplitude.

Frequencies for which the phase accumulated along a round trip in the cavity (including the random phase contribution of the rough mirror) equals an integer number of π (effective optical path corresponding to an integer number of wavelengths) will survive in the cavity while the rest will average out to nil upon successive round trips. This gives rise to a set of modes with random frequency positions, in contrast to the behaviour of FP lasers in which the phase acquired upon mirror reflection is the same for all frequencies leading to equally spaced modes determined purely by the cavity length.

The modified laser diode considered here consists of a disordered front mirror and a flat back mirror, which are modelled with random spectral responses $R_{1,2}(\nu)\times\exp[i\phi_{1,2}(\nu)]$, where $\nu$ is the frequency and $R_{1,2}(\nu)$ and $\phi_{1,2}(\nu)$ are the reflection amplitude and phase of mirror 1 (front) and 2 (back), respectively.

By introducing the mirrors' spectral response into the well-known round-trip condition[47], we obtain the lasing condition in amplitude:

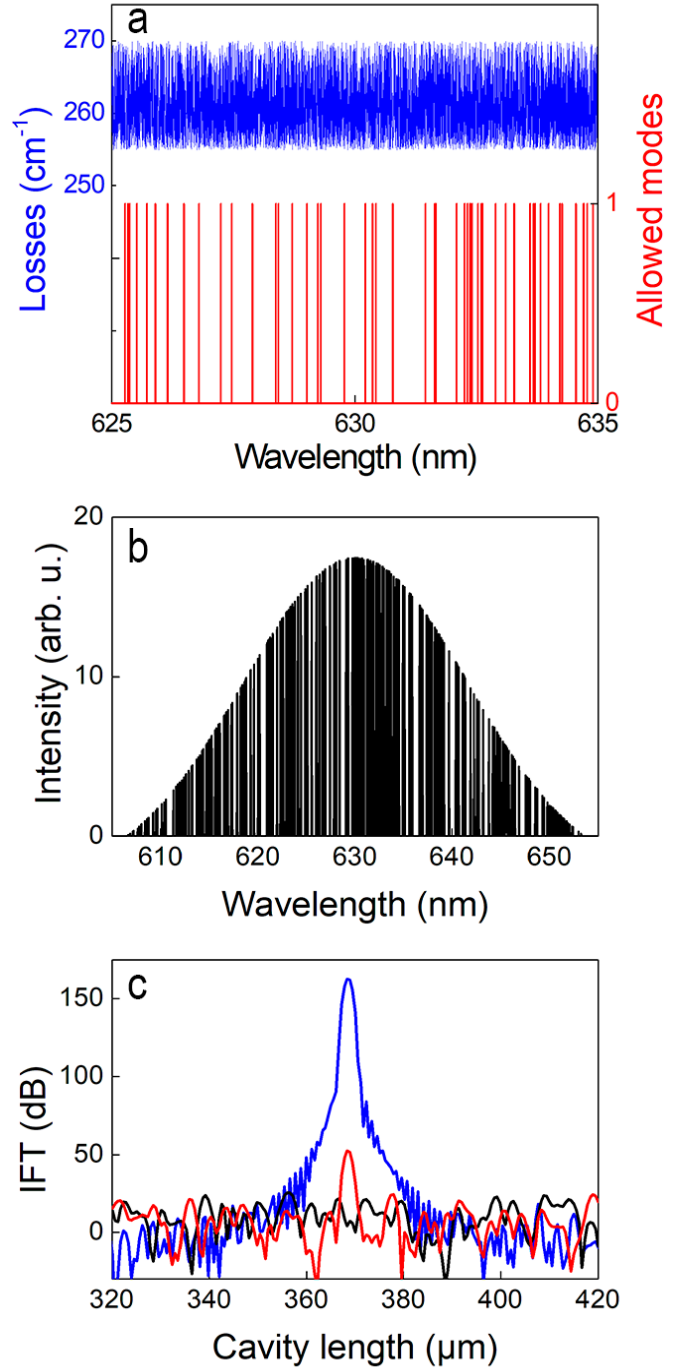

**Figure 5. | Simulation results.** (a) Simulated losses (blue line, left axis) and phase-allowed modes (red line, right axis) for the modified laser diode, (b) emission spectrum of the modified laser diode, (c) intensity Fourier transform of the emission of the original (blue line) and modified laser diode with uniform (black line) and Gaussian (red line) distributions of $R_1(\nu)$ and $\phi_1(\nu)$.

$$g_{TH}(\nu) = \alpha + \frac{1}{2L} ln\left(\frac{1}{R_1(\nu)R_2(\nu)}\right) \quad (1)$$

and phase:

$$\frac{2n\nu L}{c} + \frac{\phi_1(\nu) + \phi_2(\nu)}{2\pi} = m \quad (2)$$

where $g_{TH}(\nu)$ is the frequency dependent gain at threshold, $\alpha$ are the internal losses, $L$ is the cavity length, $n$ the refractive index and $m$ an integer number.

We numerically construct a frequency vector spanning 80 nm centred at 630 nm, with resolution 1.5 pm, and consider $L$ = 365 µm, $n$ = 3.5, and $\alpha$ = 240 cm$^{-1}$, estimated from the measured slope efficiency of the original laser diode[48]. Gain is modelled with a gaussian profile centred at 630 nm and linewidth at FWHM of 16.5 nm. More details on simulation parameters are given in Section 4 of Supplementary Material.

We numerically solve equations (1) and (2) for the original and the modified laser diode. In both cases we consider the back mirror modelled with $R_2(\nu)$ = 0.9 and $\phi_2(\nu)$ = 0, for all frequencies. For the original laser diode, the front mirror is modelled with $R_1(\nu)$ = 0.3 and $\phi_1(\nu)$ = 0, for all frequencies. For the modified laser diode, we model $R_1(\nu)$ and $\phi_1(\nu)$ with both uniform and Gaussian statistical distributions of values between 0.1 and 0.3 and between -$\pi$ and +$\pi$, respectively. The uniform distribution aims at reproducing a modified mirror with perfect randomness, while the Gaussian distribution describes an intermediate case in which the modified mirror disorder is not totally random but exhibits preferred values for $R_1(\nu)$ and $\phi_1(\nu)$.

Since losses vary randomly with frequency, due to the contribution of $R_1(\nu)$ in equation (1), the allowed modes are found at frequencies for which equation (2) is verified, as shown in Fig. 5a. Solution for the original laser diode gives constant losses with frequency and modes evenly spaced with a free spectral range equal to $2n\nu L/c$ and corresponding to 0.155 nm at 630 nm, see Section 4 of Supplementary Material.

The emission spectrum of the modified laser diode with random disorder is shown in Fig. 5b, corresponding to the net gain profile available at allowed frequency modes. For practical reasons, we limit our simulations to linear non-interacting modes, consequently no spectral narrowing is observed. A full set of equations in the framework of coupled mode theory[49] would require a time differential equation for each mode.

The IFT obtained from the spectra of the simulated original and modified laser diode are shown in Fig. 5c. A peak corresponding to a cavity length of 365 µm is observed for the original laser diode (blue line). Uniformly distributed random $R_1(\nu)$ and $\phi_1(\nu)$ correspond to pure, ideal disorder, which lead to the absence of peak in the IFT (black line). When $R_1(\nu)$ and $\phi_1(\nu)$ are modelled with Gaussian distributions, with mean value $\mu_{R1}$ = 0.2, $\mu_{\phi1}$ = 0 and standard deviation $\sigma_{R1}$ = 0.03, $\sigma_{\phi1}$ = 0.1, respectively, we still observe the peak at cavity length equal to 365 µm (red line), however it shows an intensity decrease of more than 100 dB with respect to the FP case. Therefore, the small residual periodicity observed experimentally in Fig. 4d and Fig. 4e (red line) is numerically reproduced by employing non uniform distribution for $R_1(\nu)$ and $\phi_1(\nu)$.

## Discussion

By analysing the emission properties of the modified laser diode, we identify the following five clues of multi-mode random lasing action: *i*) lasing threshold and spectral narrowing for increasing driving current, *ii*) Gaussian profile of emission with 7.8 nm FWHM above threshold with *iii*) sub-nanometre linewidth peaks at fixed frequencies exhibiting increasing amplitude as a function of injected current, *iv*) speckled intensity angular distribution and *v*) low transverse spatial coherence at any driving current.

Increase of lasing threshold (from 21 mA to 32 mA) and decrease of slope efficiency (from 0.27 W/A to 1.4 mW/A) are attributed to the roughness reduced reflectance of the front mirror and consequently increased losses. Such detrimental effect is intrinsic to the modified device, as scattering feedback is more lossy compared to the flat mirror specular reflection.

The FWHM width of the spectrum is similar in both devices below threshold, about 16 nm, and greatly differs above threshold. In the original FP laser diode, a single longitudinal mode is dominant above threshold due to mode competition sustained by the modes' long lifetime due to the high-quality cavity. On the other hand, in the modified laser diode, a Gaussian profile with FWHM width of 7.8 nm and multiple sub-nanometre spikes are observed. The occurrence of many simultaneous modes is attributed to multimode emission from modes randomly distributed in frequency, experiencing a lessen interaction, owing to reduced lifetimes, that does not allow any dominant peak to emerge.

The angular emission is characterized by an irregular speckle-like distribution of maxima and minima of intensity, opposite to the elliptical shape expected from FP laser diode. This results from the contribution of the transverse profiles of all emitting modes, due to the corrugated scattering surface.

The emergence of a large number of lasing modes also affects the spatial coherence in the modified device, resulting in low coherence values, both below and above threshold, as observed in incoherent RLs and degenerate cavities characterized by multimode emission[45].

These results are in agreement with the theoretical framework of RLs with non-distributed feedback, in which unpredictable phase contributions from disordered mirrors produce random distribution of frequency modes. In the experimentally modified laser diode, the ablation process changes the front mirror phase profile thus resulting in multimode random lasing emission.

In our vision, the evidence of random lasing is supported by the following rationale: if the modified mirror acted only as an external diffuser for the original laser diode, the speckled emission pattern would still be observed but the emission spectrum would be the same of the original laser diode. The occurrence of a completely different spectrum, with narrow randomly distributed peaks at constant wavelengths for different bias currents, is proof that the processing affects the cavity not merely adding optical elements outside the cavity that would filter or redirect the emission: the resulting device is a RL rather than a FP laser. By benchmarking the experimental results against numerical simulations, we infer that the residual periodicity observed in the experiments (the IFT peak decrease by 100 dB) is attributed to the modified mirror disorder degree. Only a complete uncorrelated random disorder would allow the elimination of the IFT peak. On the other hand, Gaussian distributions of $R_1(\nu)$ and $\phi_1(\nu)$ promote periodic modes in the spectrum, as the mean values are more recurrent than others, see details in Section 4 of Supplementary Material. Therefore, the amplitude of the IFT peak can be tuned by inducing a different degree of disorder in the modified mirror, possibly by varying the processing parameters.

## Conclusions

We demonstrated the realization of an incoherent semiconductor random laser from a commercially available FP laser diode by pulsed laser ablation of its output mirror. Our simple method requires a high energy pulsed source and a commercial laser diode, elements often found in photonics laboratories, in contrast with more complex processes required for semiconductor device fabrication, e.g. molecular beam epitaxy and/or nanoscale lithography.

We obtain a CW electrically pumped source with random multimode emission and low spatial coherence. Such a device is a practical choice for multiple applications in which optically pumped RLs are usually employed, e.g. imaging, sensing, and optical information processing.

Additionally, fine tuning pulse duration and energy provides a practical means for varying the scattering properties of the modified mirror and thus adjusting the emission characteristics, e.g. number of lasing modes, as observed in devices based on ablated dye doped polymers[40], and different degrees of spatial coherence.

With the reported results we also demonstrate the implementation of a hybrid RL device with one flat and one disordered mirror. This can open the way to the realization of more types of RLs based on non-distributed feedback and ordered/disordered structures, see Section 1 of Supplementary Material for some examples. Finally, the proposed work provides an efficient way to tailor electrically driven random lasing that will strongly promote their application as innovative light sources.

## METHODS

### Ablation

The main elements of the ablation set-up are the pulse laser source and the focusing optics, which is also used for imaging the ablated surface.

Pulse energy from the laser source (Coherent Libra, 100 fs long pulses at 800 nm with 1 kHz repetition rate) is controlled with a λ/4 waveplate and a linear polarizer. A fraction (10%) of the beam is sent via a beam splitter (BS) to a power meter connected to the PC, for monitoring the pulse energy. The number of delivered pulses is controlled via PC and a numerical routine.

Pulses are delivered by means of a 40× microscope objective (NA = 0.45). The front mirror of the laser diode is positioned out of the focusing plane of the objective, in order to shine a large area of the laser mirror.

The mirror surface is simultaneously imaged in reflection by using the same objective, a beam splitter and a white light source. Image of the illuminated mirror is projected onto a CCD camera (Pixelink PL-B776F) with a 50:50 BS and an imaging lens (focal length 5 cm). A short pass filter is used in front of the CCD camera for blocking the high energy 800 nm radiation scattered from the mirror surface.

As pulses are sent to the laser diode mirror, the ablated area is observed on the CCD camera. By moving the laser diode by means of micrometrical translational stages, we were able to place the ablation pulses on the desired position.

### Optical emission characterization

The original (Thorlabs L635P5) and modified laser diodes are fed with a current driver (Thorlabs LDC 205C) and are temperature controlled with a thermo electric cooler (Thorlabs TED 200C). CW currents and temperature of 15ºC are used in all experiments.

The output beam is collimated with a 100x microscope objective (0.9 NA) and sent to a 50:50 beam splitter.

The transmitted beam is sent to calibrated photodiode (Thorlabs S130C), connected to PC, for power measurements.

The reflected beam is sent to an imaging lens (5 cm focal length) and a fiber tip (100 μm diameter) is placed on the image plane and mounted on two computer-controlled, motorized translation stages (Thorlabs Z812). The fiber is coupled to a spectrometer (Andor Shamrock 303) and hyperspectral measurements of the far-field angular distribution emission are obtained by scanning the fiber probe on the X, Y plane over a grid of 80x80 equally spaced (100 μm) collection points.

In the set-up for the measurement of the spatial coherence of emitted radiation, the collimated output beam shines a black screen with two parallel slits with width 50 μm and placed at a distance of 500 μm. A cylindrical lens (20 cm focal length) is placed after the slits and before the CCD camera where the interference fringes are detected.

# Electrically driven random lasing from a modified Fabry-Perot laser diode

Antonio Consoli[1,2]*, Niccolò Caselli[1†#], Cefe López[1§]

# Supplementary Material

## Section 1. Non distributed feedback architectures

Lasing action of the modified laser diode presented in the main manuscript is based on non-distributed feedback random lasing. We resume here our previous results with similar architectures, describing them in the general context of laser science and proposing new possible implementations.

In Figure S1, we show our schematic vision of random lasers (RLs) with non-distributed feedback by comparing it with conventional lasing devices. Green, yellow, red and blue color code corresponds to pump, gain region, light emission and feedback elements (mirrors or scatters), respectively. We represent schematically four general classes of lasers, divided into devices with non-distributed or distributed feedback (columns) and ordered and disordered feedback elements (rows). We refer to scatters as disordered feedback elements, while mirrors and a periodic modulation of refractive index are considered as ordered feedback elements.

In this view, Fabry-Perot (FP) lasers belong to the class of non-distributed feedback devices based on ordered elements (mirrors), see Fig. S1 (a), and distributed feedback (DFB) lasers belong to the class of distributed feedback devices based on ordered elements (periodic modulation of refractive index), see Fig. S1 (b).

RLs with non-distributed feedback are obtained by simply substituting the ordered elements (mirrors) of a FP cavity with two disordered media (scatters), see Fig. S1 (c). We obtained the first demonstration of this kind of RLs with a device consisting of a liquid dye placed between two scattering volumes, i.e. agglomerations of titanium dioxide nanoparticles [1].

[1]Instituto de Ciencia de Materiales de Madrid (ICMM), Consejo Superior de Investigaciones Científicas (CSIC), Calle Sor Juana Inés de la Cruz, 3, 28049 Madrid, Spain. [2]ETSI de Telecomunicación, Universidad Rey Juan Carlos, Calle Tulipán, 28933 Madrid, Spain [†]Current address: Dep. Química Física, Universidad Complutense de Madrid, Avenida Complutense, 28040 Madrid, Spain. *e-mail: antonio.consoli@urjc.es, [#]e-mail: ncaselli@ucm.es, [§]email: c.lopez@csic.es

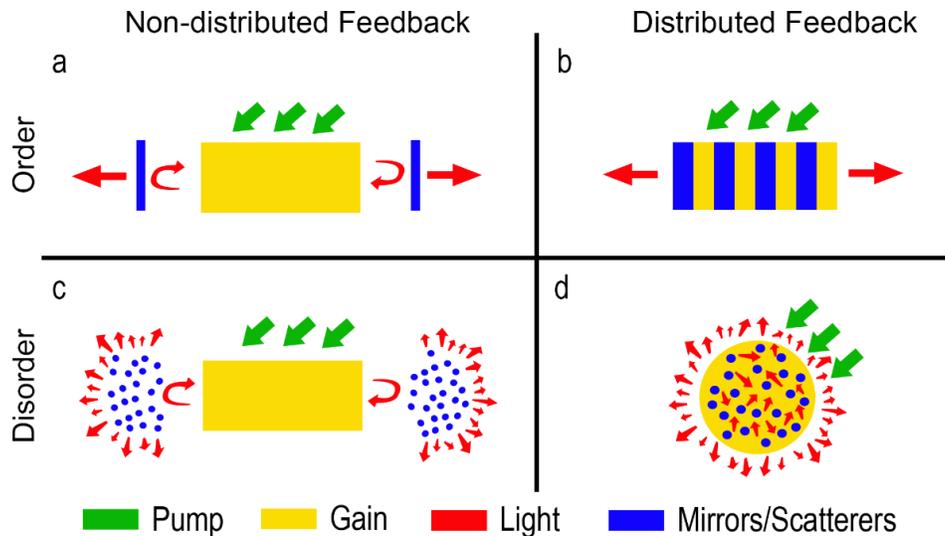

**Figure S1. | Feedback and disorder in lasers.** Fabry-Perot laser (a), DFB laser (b), RL with non-distributed (c) and with distributed feedback (d).

RLs with distributed feedback are the most diffused architecture of random lasing and consist of randomly distributed scatters inside the active medium of the device, see Fig. S1 (d). We consider that placing RLs with distributed and non-distributed feedback into this general view is a simple but relevant contribution, since this allows to draw analogies with conventional lasers, such as for example the introduction of a modified round trip condition, with frequency dependent amplitude and phase response of mirrors, for understanding why frequency modes are found at random spectral positions.

In Fig. S2, we show different alternative implementations of non-distributed feedback architectures with both ordered (mirrors) and disordered (scattering volumes or surfaces) elements.

We previously demonstrated that not only scattering volumes, as in [1], but also scattering surfaces provide feedback for random lasing, see Fig. S2 (a). This was observed in devices where scattering surface were obtained by manual cut [2], pulsed laser ablation [3] or use of natural materials [4].

In Fig. S2(b), we show the architecture used in the present manuscript, demonstrating that both ordered and disordered feedback elements can be used in the same device with non-distributed feedback, i.e. in a laser diode where a scattering surface and a mirror provide feedback for lasing.

In Fig. S2(c), we propose a structure in which a scattering volume and a mirror are used for feedback and lasing. In Fig. S2 (d) a scattering volume and a scattering surface are used.

The structures shown in Fig. S2 are examples of alternatives structures that can be obtained by mixing ordered and disordered feedback elements. Additional architectures could also be obtained by using in the same device distributed/non-distributed feedback and ordered/disordered structures.

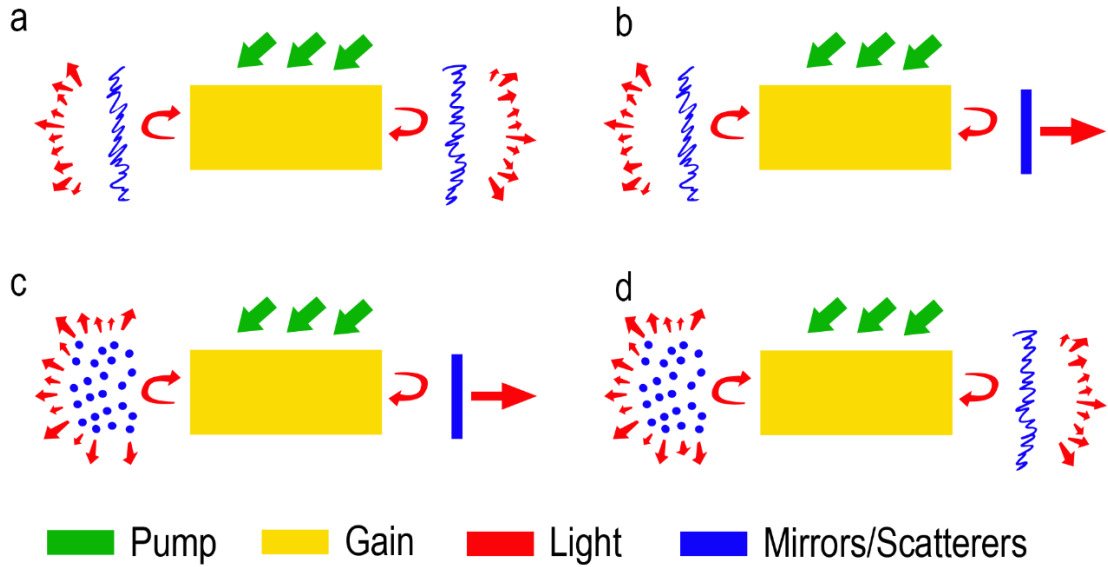

**Figure S2. | Non-distributed feedback random lasers.** RL with scattering surfaces (a) as in [2–4], RL with a scattering surface and a mirror (b) as in the present manuscript, RL with a scattering volume and a mirror (c) and RL with a scattering volume and a scattering surface (d).

## Section 2. Surface analysis of the ablated emitting area

We provide an estimation of the disorder and surface roughness obtained after pulsed laser ablation of the front FP diode mirror by analyzing the images obtained by scanning electron microscopy (SEM). The 256 gray levels of the SEM image of the emitting area are shown as in the original SEM image (a) and as a 3D surface (b). Horizontal (red line) and vertical scan directions cuts (black line) are considered in our analysis and plotted in Fig. S3 (c) and (d), respectively. Since we are interested in the roughness lateral autocorrelation, the magnitude (gray scale) representing secondary electrons intensity in the SEM image do not correspond to actual topographic heights, however it allows us to retrieve the autocorrelation length of the disordered surface scanned.

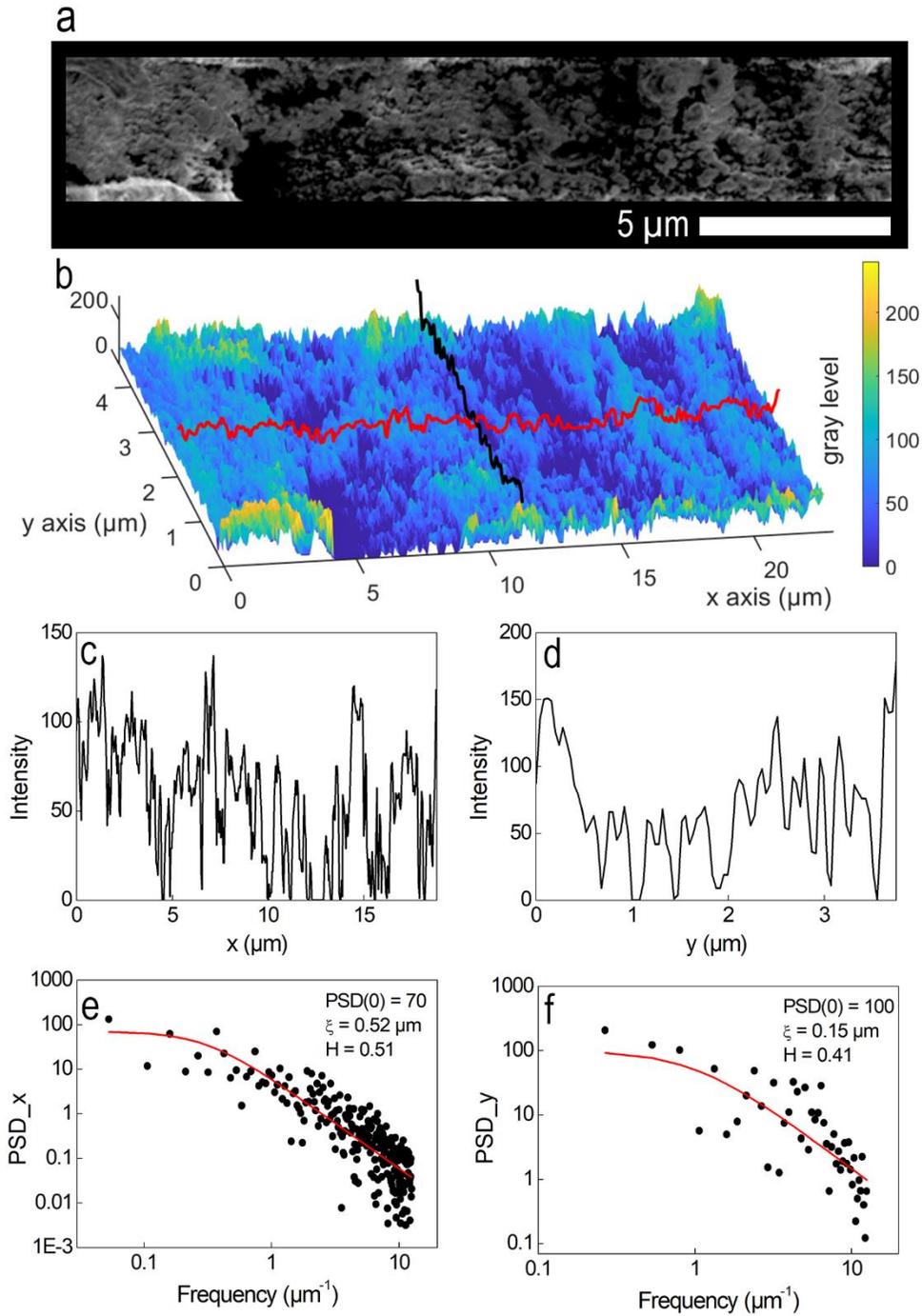

**Figure S3. | Surface characterization of the modified emitting area.** 256 gray level image of the emitting area of the modified laser diode obtained at SEM: original image (a) and 3D surface representation (b). Horizontal and vertical cross sections are highlighted by red and black lines in (b) and plotted in panels (c) and (d), respectively, the corresponding power spectral densities (PSD) are shown in (e) and (f).

The vertical axis in Fig. S3 (c) and (d) corresponds to gray levels (from 0 to 256) of the SEM image along the red and black cuts, which provide information on lateral dimensions of the roughness profile induced by the ablation process. Knowledge of the absolute depth induced by the ablation process is only partial; however since very similar materials are involved in the active region (similar III-V semiconductors) variations of secondary electron emission can be expected to be fairly faithful to the shape of the surface. Plotted profiles show that local maxima and minima are spaced at sub-micron distances and profile fluctuations are random and unpredictable.

We calculated the power spectral density (PSD) [5] of the horizontal and vertical cuts by Fast Fourier Transform (FFT) and fitted it to the equation [6]:

$$PSD(f) = \frac{PSD(0)}{1+|2\pi f \xi|^{2H+1}}$$

Where $f$ is the spatial frequency, PSD(0) is the PSD at low frequencies, $\xi$ is the correlation length in micrometers and $H$ is the Hurst exponent. Results of the fitting procedure are shown in Fig. S3 (e) and (f), for horizontal and vertical cuts, respectively.

From the fitting procedure, the correlation length $\xi$ is found to be 0.52 µm and 0.15 µm in the horizontal and vertical cross sections, respectively. This value corresponds to the distance at which two heights on a surface are no longer correlated (the autocorrelation function drops below 1/e) giving an approximate estimation of the average bump size.

Although a precise characterization of the obtained rough surface is out of the scope of our contribution, with this simple analysis we aim to approximatively quantify the spatial lateral size of the roughness induced on the surface, which we estimate in the submicron range in the order of the wavelength of light emission.

## Section 3. Far-field emission of the modified laser diode

In Fig. S4, the far-field emission [7] of the modified laser diode biased at 50 mA is shown in the wavevector space. In order to acquire this information an optical fiber with 100 µm diameter core is scanned across the back focal plane of the collecting lens so that each (X,Y) point corresponds to unique ($\vartheta,\phi$) angles of emission with respect to the original emission axis. A spectrum is acquired at each $I(x,y,\lambda) \rightarrow I(\vartheta,\phi,\lambda)$ position so that full hyperspectral characterization is obtained.

Figure S4(a) shows three points labelled as A, B and C, in Fourier space corresponding to different angles of emission. In Fig. S4 (b), spectra collected at these positions are plotted. Their lineshapes consist of randomly distributed peaks with sub-nanometer linewidths superimposed on a broad Gaussian background. At each point, the number and wavelength positions of the peaks are different, showing that each angle collects emission from different modes with different intensities thus giving rise to a specific spectral signature.

In order to illustrate the mapping of modes (i.e. in which directions is light of a given wavelength emitted) we select four narrow (0.25 nm wide) spectral regions centered at $\lambda_1$ = 629.3 nm, $\lambda_2$ = 630.6 nm, $\lambda_3$ = 632.4 nm and $\lambda_4$ = 633.2 nm (highlighted with green areas in the spectrum of Fig. S4 (b)) and extract from the $I(x,y,\lambda)$ parametric matrix four frequency resolved intensity maps $I(\vartheta,\phi\ \lambda_i)$, $i$ = 1, …, 4 of the far-field distribution, by integrating spectra in those 0.25 nm wavelength windows around the concerned $\lambda_i$.

Prior to evaluating the spectral integration, we subtracted a Gaussian background to each spectrum in order to remove the contribution of the amplified spontaneous emission that do not couple with resonant narrow modes. The Gaussian contribution was obtained from local minima of the measured spectrum found in *N* adjacent equally wide spectral regions, each value is then centered in its region and passed to a Gaussian fit function. Results shown here were evaluated for *N* = 16.

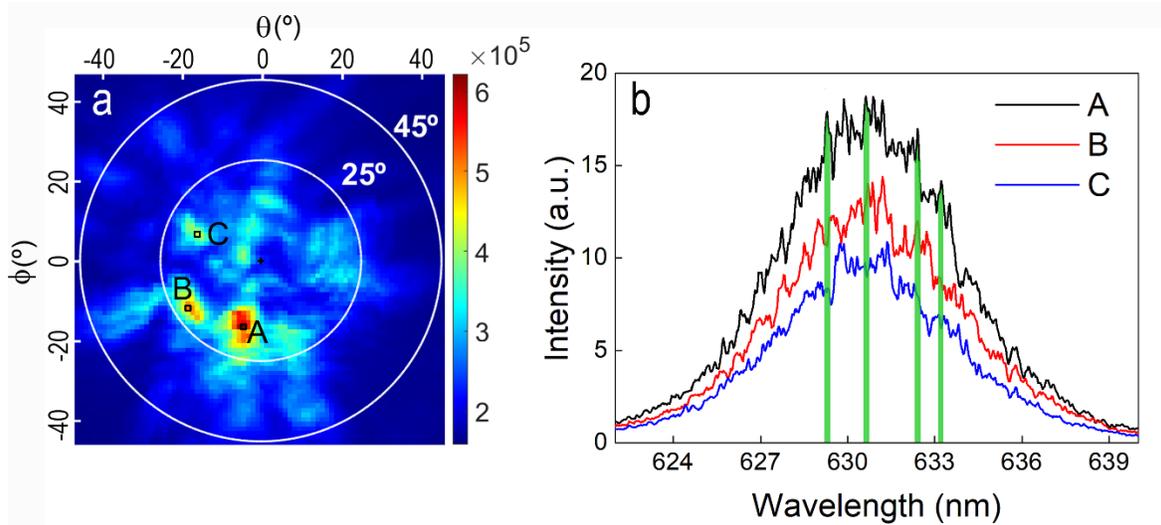

**Figure S4. | Spectrally resolved far-field emission of the modified laser diode.** (a) Intensity distribution on the far-field plane with injection current of 50 mA, obtained by integrating the spectrum collected at each point of the ($\vartheta,\phi$) plane over the entire wavelength detection range; the three black squares labelled A, B and C indicate the positions where spectra shown in panel (b) are collected. (b) Spectra collected at positions A (black curve), B (red curve) and C (blue curve): the green areas indicate the wavelength windows of integration for obtaining spectrally resolved intensity distributions centered at 629.3 nm, 630.6 nm, 632.4 nm and 633.2 nm, respectively.

If no fitting and subtraction procedure were performed, differences between maps would be hardly distinguishable, due to the large contribution of the background with respect to the relatively small intensities variations originated by the narrow peaks. We show in Fig. S5, as an example, the results of the fitting and subtraction procedure when applied to the spectrum measured in point A.

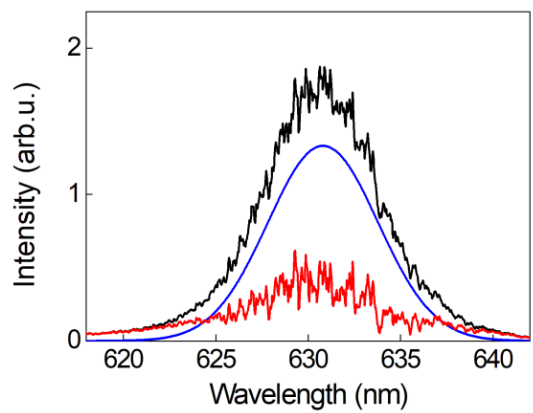

**Figure S5. | Subtraction of the gaussian background.** Measured spectrum collected at position A (black line), fitted Gaussian curve to local minima (blue line), spectrum obtained after background subtraction (red line).

In Fig. S6, we show the intensity maps obtained by integrating the collected spectra over 0.25 nm wide wavelength regions centered at 629.3 nm, 630.6 nm, 632.4 nm and 633.2 nm, after performing the background subtraction. We observe similar emission patterns which share common angles of strong emission. This behavior is attributed to a large number of lasing modes with strong spatial and spectral mutual coupling, typical of incoherent RLs.

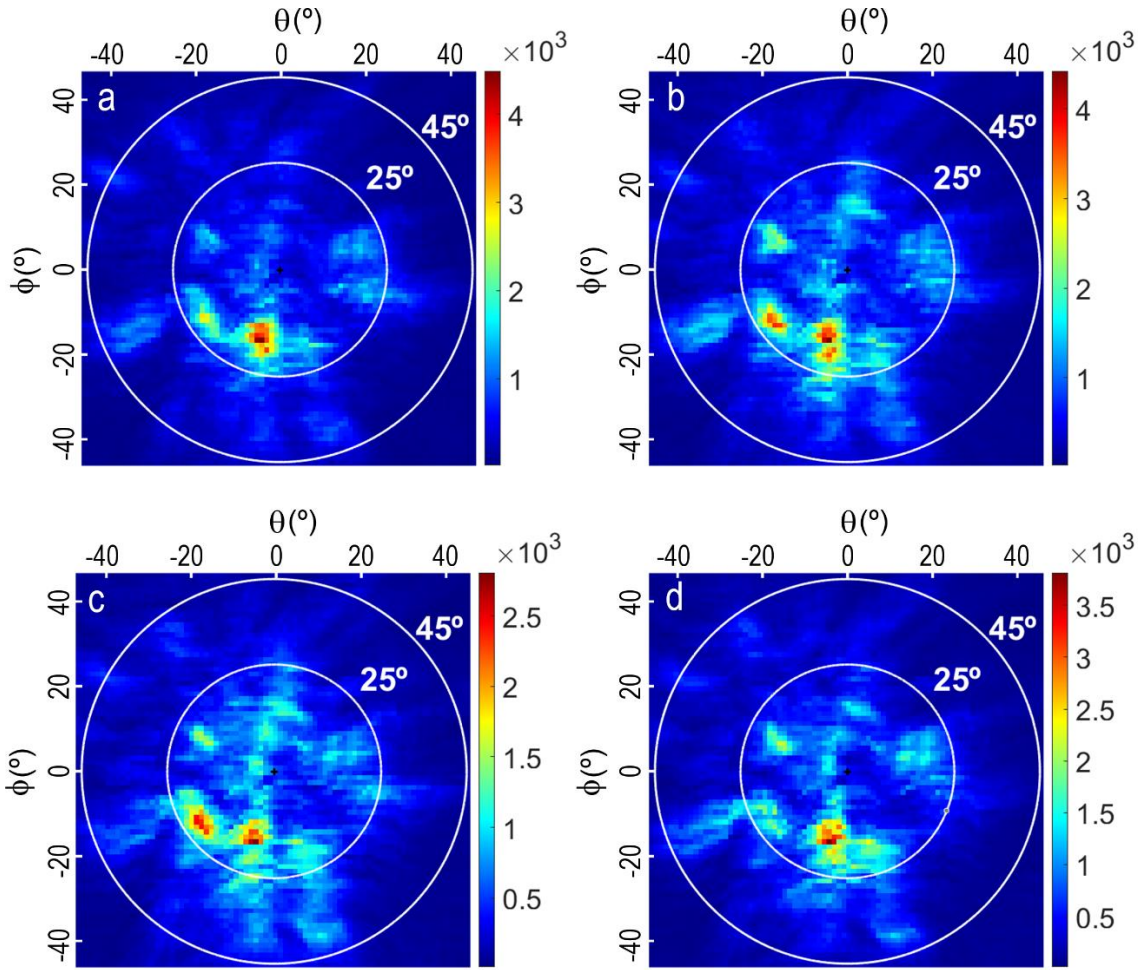

**Figure S6. | Spectrally resolved far field angular intensity distributions.** (a) Intensity distribution map obtained by integrating in 0.25 nm windows centered at 629.3 nm (a) 630.6 nm (b) 632.4 nm (c) and 633.2 nm (d).

## Section 4. Theoretical model and numerical simulations

RLs with non-distributed feedback can be treated as FP lasers with scattering mirrors, with frequency dependent amplitude and phase profiles. We introduced this simple model in [1] where the scattering media were two scattering volumes, i.e. agglomerations of $TiO_2$ nanoparticles, enclosing a liquid dye solution. In our model, a lasing mode is allowed to exist at a given frequency if a closed loop formed between the two scattering mirrors accumulates $m\pi$ optical phase taking into account the random phase contributions of scatters at that frequency. The same behavior has been observed in devices with scattering surfaces, given by the defects at the interface between air and a dye doped polymer [2,3]

In this manuscript, the modified laser diode consists of a gain medium (quantum wells) placed between the unmodified flat back mirror and the ablated rough front mirror of the original device. Accordingly we run sets of simulations for the original and modified laser diode with spectral responses $R_{1,2}(\nu)\cdot\exp(i\phi_{1,2}(\nu))$, where $\nu$ is the frequency and $R_{1,2}(\nu)$ and $\phi_{1,2}(\nu)$ are the amplitude and phase of mirror 1 (front) and 2 (back), respectively.

In all simulations, we consider for the back flat mirror $R_2(\nu)$ and $\phi_2(\nu)$ as constant with frequency and their values are fixed to: $R_2(\nu) = 0.9$ and $\phi_2(\nu) = 0$. For the original FP laser diode, $R_1(\nu) = 0.3$ and $\phi_1(\nu) = 0$, for all frequencies (we assume a lower value of reflectivity for mirror 1 as being the output mirror). For the modified laser diode, we consider both uniform and Gaussian statistical distributions for $R_1(\nu)$ and $\phi_1(\nu)$. The results obtained with uniform distributions are presented in the main manuscript, with $R_1(\nu)$ ranging between 0.1 and 0.3 and $\phi_1(\nu)$ ranging between $-\pi$ and $+\pi$, see Fig. S7 (a) and (b).

Additionally, we present here the results obtained when $R_1(\nu)$ and $\phi_1(\nu)$ have Gaussian distributions, with mean values $\mu_{R1}$ and $\mu_{\phi1}$ and standard deviations $\sigma_{R1}$ and $\sigma_{\phi1}$, respectively. In Fig. S7 (c) and (d), we show the distribution of $R_1(\nu)$, with $\mu_{R1} = 0.2$ and $\sigma_{R1} = 0.03$, and of $\phi_1(\nu)$, with $\mu_{\phi1} = 0$ and $\sigma_{\phi1} = 0.1$, respectively. These cases correspond to the Gaussian distributions employed to calculate the spectral response and its Intensity Fourier Transform (IFT) that is reported in Fig. 5 c) (red line) of the main text .

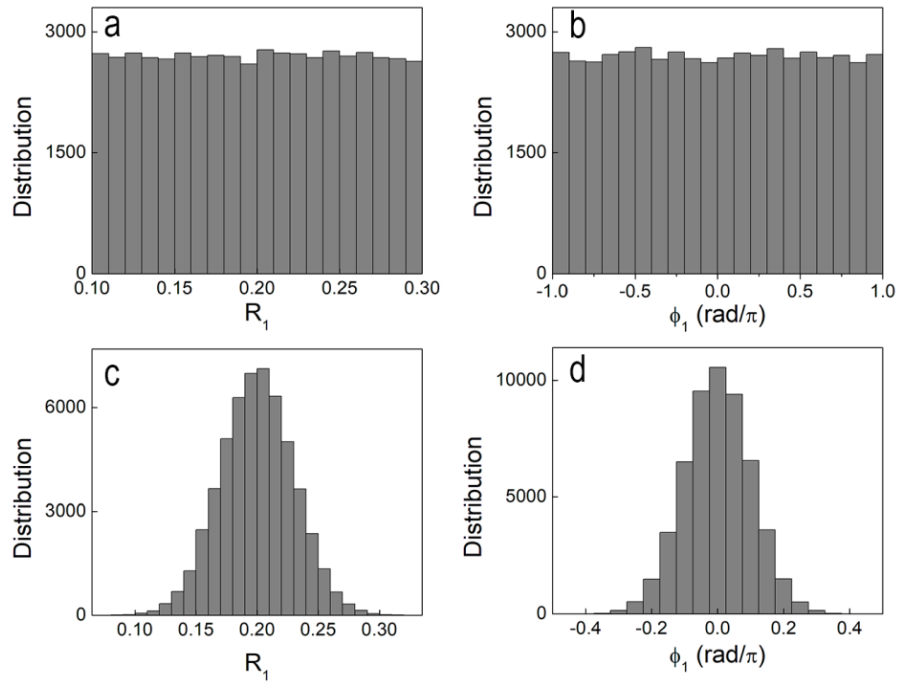

**Figure S7. | Numerically constructed spectral responses of the modified front mirror.** Amplitude (a) and phase (b) exhibiting uniform distributions, ranging between 0.1 and 0.3 and between $-\pi$ and $+\pi$, respectively. Amplitude (c) and phase (d) with Gaussian distributions, with mean values $\mu_{R1} = 0.2$ and $\mu_{\phi1} = 0$ and standard deviations $\sigma_{R1} = 0.03$ and $\sigma_{\phi1} = 0.1$, respectively.

In the computation, we construct a frequency vector ν with 54000 points, corresponding in wavelength to 80 nm range with 1.5 pm resolution centred at 630 nm. The gain is modelled with a gaussian profile centred at 630 nm and with FWHM of 16.5 nm. Losses and allowed frequencies are found for original and modified lasers, by solving the round trip equations with $R_{1,2}(\nu)\cdot\exp(i\phi_{1,2}(\nu))$:

$$g_{TH}(\nu) = \alpha + \frac{1}{2L}\ln\left(\frac{1}{R_1(\nu)R_2(\nu)}\right) \quad (1)$$

$$\frac{2n\nu L}{c} + \frac{\phi_1(\nu)+\phi_2(\nu)}{2\pi} = m \quad (2)$$

where $g_{TH}(\nu)$ is the frequency dependent lasing threshold, $\alpha$ are the internal losses, L = 365 μm is the cavity length, n = 3.5 the refractive index, c the speed of light and m an integer number. Internal losses, $\alpha$ = 240 cm$^{-1}$, were estimated from the measured slope efficiency, $\eta_{FP}$, of the original laser diode [8], considering that $\eta_{FP} = (\alpha_{mir}/(\alpha_{mir}+\alpha))\cdot h\nu_0/q$, where $\alpha_{mir} = 0.5\cdot L^{-1}\cdot\ln[(R_1(\nu)\cdot R_2(\nu))^{-1}]$, with $R_1(\nu)$ = 0.3 and $R_2(\nu)$ = 0.9 for all ν, h is the Planck's constant, $\nu_0$ the central frequency and q the electron charge.

Simulation results for the original laser diode are shown in Fig. S8 . The simulated original laser diode shows constant losses of 255 cm$^{-1}$, given by the right side of Eq. (1), which corresponds to the sum of internal and constant mirror losses. By solving Eq. (2), we obtain equally spaced modes with a free spectral range of 0.155 nm, corresponding to the FP cavity length L = 365 μm.

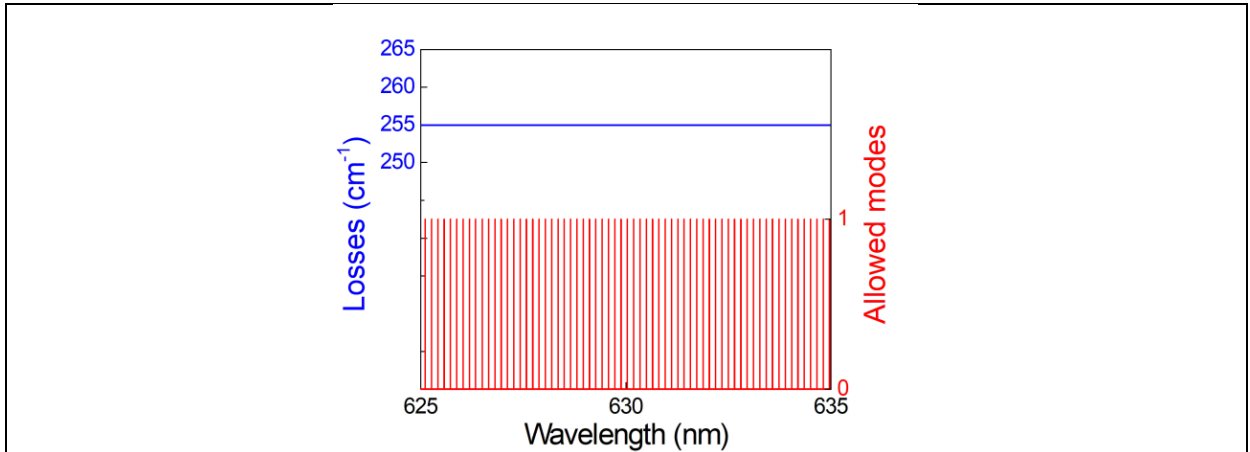

**Figure S8. | Simulation results: original FP laser diode.** Losses (blue line curve) and allowed modes (red line curve) of the simulated original laser diode.

Results obtained by employing uniform distributions of $R_1(\nu)$ and $\varphi_1(\nu)$ are presented in the main text, in Fig. 5 (a)-(b) and in Fig. 5 (c) black line. Here, results obtained with Gaussian distributions of $R_1(\nu)$ and $\phi_1(\nu)$ are presented in Fig. S9. Losses vary with frequency, with an average value of 260 cm$^{-1}$, due to the contribution of $R_1(\nu)$ in $\alpha_{mir}$, see the blue line curve in Fig. S9. Allowed modes, plotted with red line curve in Fig. S9, are obtained at frequencies where Eq. (2) is verified., resulting in the emission spectrum shown by the black line in Fig. S9.

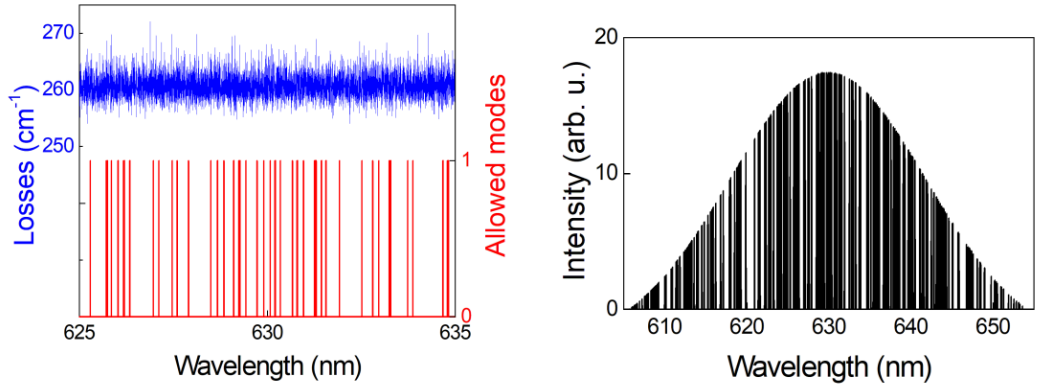

**Figure S9. | Simulation results: modified laser diode with reflectivity and phase exhibiting Gaussian distribution.** Losses (blue line), allowed modes (red line) and spectral response (black line) of the modified laser diode.